\documentclass[aps,prl,twocolumn,reprint,showpacs,superscriptaddress,footinbib,groupedaddress]{revtex4-1}
\usepackage{dsfont}
\usepackage{amsmath}
\usepackage{graphicx}
\usepackage[utf8]{inputenc}
\usepackage{ulem} %added to be able to cross text out
\usepackage{units} % for using nicefrac in texts
\usepackage{version} % for comment
\usepackage{color}
\usepackage{colortbl}
\usepackage{xcolor}  % for writting parts in color, e.g. in red
\usepackage{fancyref}
\usepackage[colorlinks=true,citecolor=black,urlcolor=blue,linkcolor=black]{hyperref} % choice of color for hyperref links

\usepackage{multirow}
\usepackage{array}
\usepackage{booktabs}
\usepackage{longtable}
\newcommand{\ata}{\mathcal{A}}
\newcommand{\atb}{\mathcal{B}}

\newcommand{\eff}{_{\text{tot}}}
\newcommand\stxt[1]{_{\text{#1}}} % subscript
\newcommand{\ms}{\ \text{ms}}

\newcommand{\mrad}{\ \text{mrad}}

\newcommand{\Rba}{$^{85}$Rb}
\newcommand{\Rbb}{$^{87}$Rb}

\newcommand{\beq}{\begin{equation}}
\newcommand{\eeq}{\end{equation}}

\begin{document}
\graphicspath{{Figures/}}
\preprint{AIP/123-QED}

\title{Proposal for a Quantum Test of the Weak Equivalence Principle with Entangled Atomic Species}

\author{Remi Geiger}
\email{remi.geiger@obspm.fr}
\affiliation{LNE-SYRTE, Observatoire de Paris, Sorbonne Universit\'e, PSL Universit\'e Paris, CNRS, 61 avenue de l'Observatoire, 75014 Paris, France.}

\author{Michael Trupke}
\email{michael.trupke@univie.ac.at}
\affiliation{Vienna Center for Quantum science and technology (VCQ), Faculty of Physics,  Research Platform TURIS, University of Vienna, Boltzmanngasse 5, A-1090 Vienna, Austria}

\date{\today}

\begin{abstract}
\label{abstract}
We propose an experiment to test the Weak Equivalence Principle (WEP) with a test mass consisting of two  entangled atoms of different species. 
In the proposed experiment, a coherent measurement of the differential gravity acceleration between the two atomic species is considered, by entangling two atom interferometers operating on the two species. 
The entanglement between the two atoms is heralded at the initial beam splitter of the  interferometers through the detection of a single photon emitted by either of the atoms, together with the impossibility of distinguishing which atom emitted the photon.  
In contrast to current and proposed tests of the WEP, our proposal explores the validity of the WEP in a regime where the two particles involved in the differential gravity acceleration measurement are not classically independent, but entangled.
We propose an experimental implementation using \Rba \ and \Rbb \ atoms entangled by a vacuum stimulated rapid adiabatic passage protocol implemented in a high finesse optical cavity. We show that an accuracy below $10^{-7}$ on the E\"otv\"os parameter can be achieved.

\end{abstract}

% for PACS : http://www.aip.org/publishing/pacs/pacs-alphabetical-index
% 06.30.Ft, 95.55.Sh = clocks
% 37.25.+k, 03.75.Dg = atom interferometry

%\keywords{}

\maketitle

\label{introduction}

%\paragraph{Introduction.} 
The current understanding of gravity is formulated by the theory of General Relativity  which has been proven to accurately describe many astronomical phenomena.  
The Weak Equivalence Principle (WEP), also known as the Universality of Free Fall, represents one of the three pillars of the Einstein Equivalence Principle, which was at the basis of the elaboration of General Relativity \cite{Will2006}. 
According to Damour \cite{Damour2012}, the Equivalence `Principle' is  not satisfactory, as it sets an absolute structure for fundamental coupling constants (e.g. the fine structure constant), in contrast to how physics (and relativity in particular) is constructed, i.e. avoiding the assumption of absolute structures.  
Unification theories, which aim at describing gravity and the three interactions of the Standard Model within a single mathematical framework, therefore commonly imply violations of the Equivalence Principle. WEP tests thus represent key probes in the search of new physical phenomena \cite{Damour2012}.
As the types of WEP violations, as well as the levels at which they could occur, are theoretically elusive, an experiment with  improved accuracy or involving a different type of test mass might therefore point towards new physics \cite{Damour2012}. 

WEP tests are quantified by the E\"otv\"os parameter $\eta=2(a_A-a_B)/(a_A+a_B)$, which deviates from zero if the accelerations $a_A$ and $a_B$ of the two bodies are different in a given gravitational field.
 WEP has been tested at the level of $10^{-13}$ uncertainty on the E\"otv\"os parameter in continuously improved experiments involving torsion balances \cite{Wagner2012} or Lunar Laser Ranging \cite{Williams2004}. 
The   MICROSCOPE experiment, currently in Earth orbit, aims at improving such tests  to the level of $10^{-15}$  by using two free-falling macroscopic differential accelerometers. The first results  recently published in Ref.~\cite{Touboul2017} show the validity of WEP at the level of $2\times 10^{-14}$.
 Apart from these high precision experiments involving macroscopic  masses, efforts are also being pursed to test the WEP with microscopic or exotic  particles.
These efforts started with experiments involving electrons  \cite{Witteborn1967} and neutron interferometers \cite{Colella1975,Bonse1983,Abele2012}.
More recently, several results  with  cold atoms have been reported \cite{Fray2004,Tarallo2014,Schlippert2014,Zhou2015,Bonnin2015,Duan2016,Rosi2017}, together with  proposals for improved tests  \cite{Dimopoulos2007,Varoquaux2009,Altschul2015}.
Experiments using antimatter are also being developed \cite{Debu2012,Kellerbauer2008}.

The  WEP and the role of inertial and gravitational masses in quantum mechanics have been studied theoretically in numerous works, see e.g. \cite{Viola1997,Kajari2010}. 
%For example, Ref.~\cite{Viola1997} discusses the consequence of WEP on the time of flight probability distributions of superposition states, and Ref.~\cite{Kajari2010} analyzes the role of inertial and gravitational masses in quantum mechanics.
%A space experiment has been proposed to study the effect of gravity on entanglement between excitations of  Bose–Einstein condensates undergoing a change in gravity field \cite{Bruschi2014}.
It was shown recently in Ref.~\cite{Zych2015} that the validity of the Equivalence Principle for classical objects does not imply the validity of its quantum formulation, i.e. the equivalence between inertial and gravitational mass operators. Such considerations point towards new experimental approaches involving quantum test particles described by superposition states of internal degrees of freedom, e.g. as  proposed in \cite{Orlando2016}. Very recently, an atom interferometry test of such a quantum formulation of the Equivalence Principle has been performed, by measuring the free-fall acceleration of an atom in a superposition of different internal energy states  \cite{Rosi2017}.

In this letter we propose a test of the WEP with a fundamentally different type of object than in previous or ongoing experiments, namely two entangled atoms of different species.
The experiment considers the comparison of the free-fall acceleration of an atom $\ata$ when it is entangled with a different atomic species $\atb$, to the free-fall acceleration of the atoms without entanglement. We describe a particular implementation with \Rba \ and \Rbb \ atoms, and an entangling  process based on a vacuum stimulated rapid adiabatic passage protocol implemented in a high finesse optical cavity.  

%\paragraph{Principle.}
\label{par:principle}
The concept of our proposal relies on a vertical atom interferometer in which atomic species $\ata$ and $\atb$ are entangled. The entanglement is heralded at the first beam splitter of the interferometer by the detection of a single photon. The scheme is related to the seminal work in Refs.~\cite{Cabrillo1999,Slodicka2013}, but operates here on freely propagating, distinguishable atoms instead of trapped, identical particles. 
In the event of the emission of a single photon from one of the two atoms in the direction of a photon detector, and assuming that it is not possible to distinguish which atom emitted the photon, a detection event will herald a superposition state: Atom $\ata$ acquires the  momentum $\hbar\vec{k}$ ($\ata$ emitted the photon of wavevector $\vec{k}$) and atom $\atb$ is left unperturbed, or vice-versa. The corresponding entangled state can be written as:
\beq
|\psi\rangle = \frac{1}{\sqrt{2}} \left(|\ata,\hbar\vec{k}; \atb,\vec{0}\rangle + e^{i\phi}|\ata,\vec{0}; \atb,\hbar\vec{k}\rangle \right).
\label{eq:superposition}
\eeq
The beam splitter thus creates a superposition of the momenta of the two atomic species $\ata$ and $\atb$, with $\phi$ a fixed (non-random) phase in the case of a coherent superposition.
To complete the interferometer, the two paths produced at the first beam splitter are subsequently manipulated with conventional atom optics (e.g. two-photon Raman transitions \cite{Kasevich1991}) in order for the paths of each species to interfere. 
Single atom detectors are finally used to probe the atomic interference at the interferometer output.

\begin{figure}[!h]
\includegraphics[width=\linewidth]{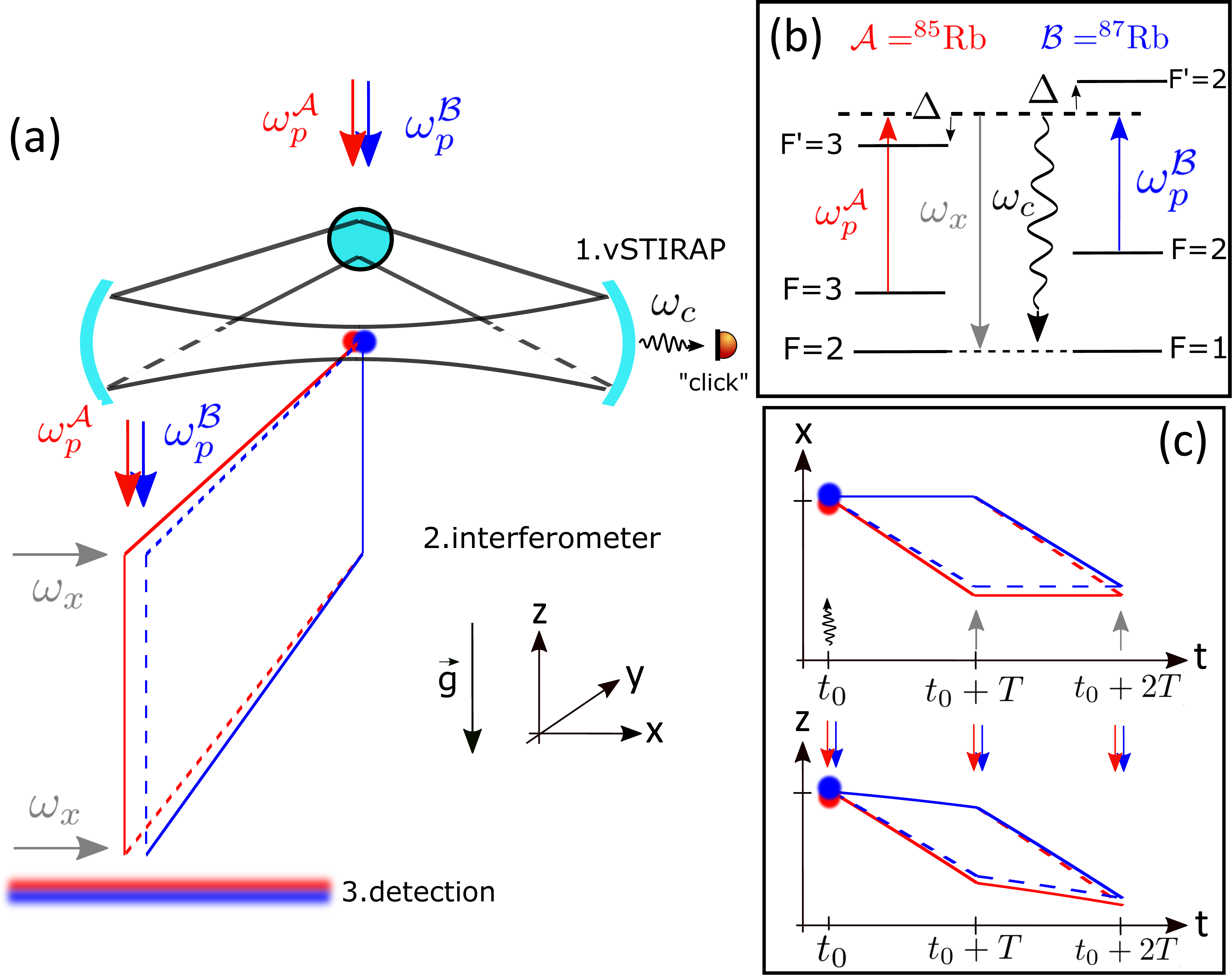}
\caption{\label{fig:implementation} 
Implementation of the entangled interferometers with \Rba \ and  \Rbb \  atoms and a vSTIRAP protocol to realize the entangling beam splitter. 
(a) General sketch of the  experiment: after laser cooling, the atoms are released  and let to fall under gravity in a high finesse optical cavity made of three mirrors lying in the $(xy)$ plane. 
During the vSTIRAP process, a photon is extracted from the pump beam (red and blue arrows for \Rba \ and \Rbb, respectively), and a  photon is emitted into the cavity mode. The emitted photon (frequency $\omega_c$) is detected at one output of the cavity (`click').
% Prior to the entrance in the cavity, the atoms are filtered to enter the cavity mode and interferometer with the correct velocity distribution and internal energy state.
(b) Energy levels of the atoms subject to two-photon Raman transitions. The high-finesse cavity is resonant for a mode of frequency $\omega_c$. The vSTIRAP process is initiated at time $t=t_0$ by a pulse of the pump beams of frequency $\omega_p^{\ata,\atb}$. 
The gray arrow represents a laser beam (frequency $\omega_x=\omega_c$) used to perform the Raman transitions in the mirror pulse (at $t=t_0+T$) and final beam splitter pulse (at $t=t_0+2T$) of the interferometer.
(c) Space-time diagrams of the atom interferometer in the $x$ and $z$ directions. Note the two-dimensional feature of the atom optics, which transfer momentum along the $x$ (cavity beam and gray arrow) and $z$ (red/blue arrows) directions.
 In panels (a) and (c), the difference in recoil velocities between  \Rba\ and \Rbb \  has been exaggerated to $10\%$ (instead of $2.3\%$). In the bottom of panel (c), gravity has been reduced to $g=0.01 \ \text{m.s}^{-2}$ in order to highlight the recoil effect.
}
\end{figure}

%\paragraph{Implementation.}
\label{par:implementation}
We  focus in this letter on a particular implementation of this idea using \Rba \ and  \Rbb \ atoms, as sketched in Fig.~\ref{fig:implementation}. To entangle the two atoms, we propose to employ a vacuum stimulated Raman adiabatic passage (vSTIRAP) protocol \cite{Hennrich2000}, where the detection of a single photon exiting a high-finesse optical ring cavity heralds the entangled state of Eq.~\eqref{eq:superposition}. 
The cavity is on resonance with a mode of frequency $\omega_c$. 
The two atoms are initialized in one of their two hyperfine ground states, respectively $|F=3\rangle$ for $\ata=$\Rba \ and $|F=2\rangle$ for $\atb=$\Rbb, see Fig.~\ref{fig:implementation}(b). 
The vSTIRAP process is triggered at time $t=t_0$ by a pulse of two pump laser beams at frequencies $\omega_p^{\ata}$ and $\omega_p^{\atb}$ (red and blue vertical arrows), which  fulfil the two-photon Raman resonance condition for each atom: $\omega_p^{\alpha}-\omega_c = G^{\alpha} + \omega_R^{\alpha}$, where $G^{\alpha}$ is the hyperfine  splitting frequency, and $\omega_R^{\alpha}$ is the two-photon recoil frequency, with $\alpha=\ata,\atb =$\ \Rba,\ \Rbb  \ \cite{footnoteDoppler}. 
 Assuming that the probability of the adiabatic passage for each atom is small \cite{Cabrillo1999,Slodicka2013}, the vSTIRAP process will in all likelihood deposit at most a single photon into the cavity. The photon can then escape the cavity while one of the atoms is transferred from one hyperfine state to the other \cite{Hennrich2000,Wilk2007}. If the photon emission of both atomic species can be made to have the same envelope and frequency, then a detection event will herald the desired entangled state. 

In view of the WEP test, we aim to measure the gravitational acceleration with the atom interferometer, requiring a vertical accelerometer \cite{Kasevich1991}. Therefore, at least one of the light beams realizing the Raman transition must have a projection on the gravity direction ($z$). 
We choose a configuration where the cavity is horizontal ($xy$ plane in Fig.~\ref{fig:implementation}) and where the pump beams are aligned with gravity. 
As a consequence, the beam splitter operates in two dimensions, with a transfer of momentum $\hbar \vec{k}\eff\equiv \hbar( k_x \hat{x}-k_z\hat{z})$ along the $\hat{x}$ and $\hat{z}$ direction, with $k_x=\omega_c/c$  (resp. $k_z$) the wavevector of the cavity (resp. pump) photon.
The remaining part of the interferometer is a typical Mach-Zehnder  configuration  \cite{Kasevich1991}, apart from the fact that the mirror and final beam splitter pulses are two-dimensional in the momentum transfer, see Fig.\ref{fig:implementation}(c). 
%In particular, the wavepackets associated to the two paths are  separated in the $\hat{x}$ and in the $\hat{z}$ directions  at the mirror pulse occurring at time $t=t_0+T$.

After the last beam splitter pulse occurring at time $t=t_0+2T$, the detection of each single-atom state can be performed by fluorescence detection with a photodiode \cite{Parazzoli2012}, or by imaging using a light sheet detector \cite{Buecker2009}. 
%In that case, two light sheets are required in order to identify each atom (ideally four light sheets to probe the presence of each atom in each of the two hyperfine states).  
%In the next paragraphs, we will compute the interferometer phase shifts, and give details on the main technical requirements of the experiment.

\paragraph{Interferometer phase shift.}
\label{par:interferometer_phase}
We compute the atom interferometer phase shift following the  path integral approach \cite{Storey1994}. 
In  atom interferometers using two-photon Raman transitions, the phase of the interferometer originates from the relative   phase  between the Raman lasers, $\phi(t)$, which is imprinted on the diffracted atomic wave by the different Raman pulses \cite{Borde2004GRG,Wolf2011}. 
More precisely, the  phase shift imprinted on atom $\alpha=\ata,\atb$ by a light pulse is  $\phi_{\alpha}(t)= \vec{k}\eff^{\alpha}\cdot\vec{r}^{\alpha}(t) + \varphi_0^{\alpha}(t)$, with  $\vec{r}^{\alpha}(t)$  the position of the atom in the laboratory frame holding the lasers and the cavity, and $\varphi_0^{\alpha}(t)$  a   phase offset  associated with the change of the internal energy state.
%The finite duration of the light pulse can be neglected if the relative phase between the two lasers is stabilized between all pulses. 
Assuming that all Raman lasers are phase-locked (i.e. red/gray and blue/gray lasers in Fig.~\ref{fig:implementation}), we can leave aside the $\varphi_0$ term, and neglect the finite duration of the Raman pulse ($\sim 10 \ \mu$s typically). The laser phase can then be written more explicitly as $\phi^{\alpha}(t)=-k_x x_{\alpha}(t)- k_z^{\alpha} z_{\alpha}(t)$. Note that $k_x$ is the same for both atoms (gray arrow); the relative difference in $k_z$ is $\sim 10^{-5}$ (difference in hyperfine splitting between \Rba \ and \Rbb) and will be omitted from now on \cite{SupplMat}.

After the vSTIRAP process, the two-particle state reads
\beq
|\psi(t_0)\rangle = \frac{1}{\sqrt{2}} \left(|\ata,\hbar\vec{k}\eff; \atb,\vec{0}\rangle e^{i\phi^{\ata}_0} + |\ata,\vec{0}; \atb,\hbar\vec{k}\eff\rangle e^{i\phi^{\atb}_0} \right),
\label{eq:state_after_BS}
\eeq
with $\phi^{\alpha}_0\equiv\phi^{\alpha}(t_0)$. Note that we have treated the phase shift imprinted on the atom during the vSTIRAP process as for a conventional Raman transition, although the emission of the photon occurs in the vacuum of the cavity mode \cite{Chapman1995}.
%We neglect the finite duration ($\sim$ few $\mu$s typically) of the vSTIRAP pulse.
In the Raman process, the change of momentum $\vec{0}\leftrightarrow \vec{k}\eff$ is accompanied by a change of the hyperfine state  of the atom \cite{Borde1989}, which we omit in Eq.\eqref{eq:state_after_BS} to simplify the notations. 

After the mirror pulse at time $t_0+T$, the state reads
\begin{eqnarray}
|\psi(t_0+T)\rangle & = & \frac{1}{\sqrt{2}} \Big[|\ata,\vec{0}; \atb,\hbar\vec{k}\eff\rangle e^{i\phi^{\ata}_0} e^{i(\phi^{\atb}_T-\phi^{\ata}_T)}   \\ \nonumber
& + & |\ata,\hbar\vec{k}\eff; \atb,\vec{0}\rangle e^{i\phi^{\atb}_0} e^{i(\phi^{\ata}_T-\phi^{\atb}_T)}   \Big],
\label{eq:state_after_mirror}
\end{eqnarray}
with $\phi_T^{\alpha}\equiv \phi^{\alpha}(t_0+T)$ the relative Raman laser phase at time $t_0+T$.  
The last beam splitter occuring at $t_0+2T$ acts globally on both atoms \cite{Slodicka2013}, which results in the output state
\begin{eqnarray}
\nonumber
|\psi(t_0+2T)\rangle & = & \frac{1}{2\sqrt{2}} 
\Big[|\ata,\vec{0}; \atb,\vec{0}\rangle \left( i e^{i(\varphi-\phi^{\ata}_{2T})} + i e^{i(\Psi - \phi^{\atb}_{2T})} \right)   \\ \nonumber
& + & |\ata,\hbar\vec{k}\eff; \atb,\hbar\vec{k}\eff \rangle \left( i e^{i(\varphi+\phi^{\atb}_{2T})} + i e^{i(\Psi+\phi^{\ata}_{2T})}\right)    \\ \nonumber
& + & |\ata,\vec{0}; \atb,\hbar\vec{k}\eff \rangle \left( i^2   e^{i(\varphi-\phi^{\ata}_{2T}+\phi^{\atb}_{2T})} +  e^{i\Psi}\right)    \\ 
& + & |\ata,\hbar\vec{k}\eff; \atb,\vec{0} \rangle \left(   e^{i\varphi} +i^2  e^{i(\Psi+\phi^{\ata}_{2T}-\phi^{\atb}_{2T})}\right)    
 \Big]
\label{eq:output_state}
\end{eqnarray}
where $\varphi = \phi^{\atb}_0 +\phi^{\ata}_T-\phi^{\atb}_T  $ and $\Psi = \phi^{\ata}_0 +\phi^{\atb}_T-\phi^{\ata}_T  $.

The detection of the four possible states at the interferometer output can be performed by fluorescence detection (light sheets in Fig.~\ref{fig:implementation}), which resolves the two hyperfine states of each atom \cite{Borde1989}. For example, the probability of detecting atom $\ata$ and atom $\atb$  in the output port corresponding to the null momentum (projector on state $|\ata,\vec{0}; \atb,\vec{0}\rangle $) is given by  
\beq
P_{00} = \left | \langle \ata,\vec{0}; \atb,\vec{0}  |\psi(t_0+2T) \rangle \right |^2 = \frac{1}{8} \left |1+e^{i(\Phi_{\ata}-\Phi_{\atb})} \right |^2, 
\eeq
with $\Phi_\alpha = \phi^\alpha_0-2\phi^\alpha_T+\phi^\alpha_{2T}$. 

The expression of the phase shift  $\Phi_\alpha$  is the same as in  a traditional three light pulse interferometer \cite{Storey1994}.
However, in contrast to two classically independent interferometers that would operate in parallel on atom $\ata$ and atom $\atb$, the phase of the entangled interferometer, $\Delta\Phi\equiv\Phi_{\ata}-\Phi_{\atb}$,  is determined by the phase shifts experienced by both atoms, as a result of two-particle interferometry \cite{Horne1989,Rarity1990}. 
The entanglement between the two interferometers can thus be verified experimentally by applying controlled  phase shifts on the relative phase of the (phase-locked) Raman lasers: while a phase shift applied to only one pair of lasers (say for $\ata$) affects the mutual signal $P_{00}$, the same phase shift applied on both pairs of lasers should not affect $P_{00}$.

Finally, $\Delta\Phi$ results from the terms in Eq.\eqref{eq:output_state}, and writing the trajectories of the atoms as $x^\alpha(t)=x^\alpha_0 + v^\alpha_{x0}(t-t_0)+a_x^{\alpha}(t-t_0)^2/2$ and $z^\alpha(t)=z^\alpha_0 + v^\alpha_{z0}(t-t_0) - g_z^{\alpha}(t-t_0)^2/2$ we obtain:
\beq
\Delta\Phi =   k_z( g_z^\ata - g_z^\atb) T^2 + k_x( a_x^\ata -   a_x^\atb) T^2,
\label{eq:final_phi}
\eeq 
which reflects the bidirectional acceleration sensitivity of the interferometer.
Provided that the experiment is not constantly accelerated in the horizontal direction with respect to the freely falling atoms ($a_x^\alpha=0$), the second term vanishes on average.
%We also neglect the Sagnac phase shift  associated with the physical area of the interferometer, as it could be independently extracted by reversing the Raman wavevector \cite{Dutta2016}.
The main phase shift of the interferometer, $\Delta\Phi\stxt{WEP}\equiv k_z( g_z^\ata -   g_z^\atb) T^2$, represents a coherent measurement of the difference in the gravitational acceleration between the two atoms.

\paragraph{Details of implementation and expected sensitivity.}
\label{par:requirements}
The design of the experiment is fundamentally driven by the need for indistinguishability of the emitted photon during the vSTIRAP process, and the indistinguishability of the two atoms in the interferometer up to the last beam splitter.
On the technical aspects, the design must take into account \textit{(i)} the preparation of two cold  atoms of \Rba \ and \Rbb \ with
high probability; \textit{(ii)} the design of the high finesse ring cavity; \textit{(iii)} the optical access for the laser beams realizing the mirror and final beam splitter pulses; and \textit{(iv)} the detection of the two atoms. 
We  consider that the atoms are loaded directly in the cavity mode, and exit the cavity for the second and third interferometer pulses, which requires a sufficient interrogation time $T$. We consider $T=50 \ms$ in the following.

%We also consider that the atoms exit the cavity for the second and third interferometer pulses, which might require some free fall time $t_0\sim 100 \ms$ before the vSTIRAP pulse (not shown in Fig.~\ref{fig:implementation}). 

%A first source of distinguishability originates from the phase of the emitted photon which might randomly vary because of the random position of the atoms associated with their finite temperature \cite{Wilk2010,Slodicka2013}. Considering that the two atoms fall at different positions in the cavity mode in the $x$ direction, the variance of the phase of the photon is $\langle\Delta\phi^2\rangle\sim (k_x t\stxt{TOF} \sigma_v)^2$ where $t\stxt{TOF}$  is the time of flight between the cold atom source and the cavity mode and $\sigma_v$ is the rms velocity of the cold atoms. Preserving a fidelity  $\mathcal{F}=\langle \Psi|\hat{\rho}|\Psi\rangle \simeq e^{-\langle\Delta\phi^2\rangle/2}$ (with $\hat{\rho}$ the two-atom density matrix \cite{Wilk2010}) as close as one as possible implies  $\sigma_v\ll (k_x t\stxt{TOF})^{-1}$. Following \cite{Wilk2007}, a falling distance of few cm between the cold atom source and the cavity mode implies $t\stxt{TOF}\sim 100 \ms$, resulting in $\sigma_v\ll 1 \ \mu\text{m.s}^{-1}$. DISCUSSION OF FILTERING TECHNIQUES (controllable pinhole with light shifts, or mechanical pinhole of $\sim \ \mu m$ size).

The first step consists in preparing two cold ($\sim$ few $\mu$K) \Rba \ and \Rbb \ atoms, which can be achieved in microscopic dipole traps operating in the collisional blockade regime \cite{Schlosser2002,Parazzoli2012,Xu2015}. 
The \Rba \ and \Rbb \ atoms are prepared before release into the cavity in the states $|F=3, m_F=3\rangle$ and $|F=2, m_F=2\rangle$, respectively, and driven with individual, $\pi$-polarized pump beams. 
We envisage a ring cavity with coupling strength, field amplitude decay and atomic decay rates $\lbrace g,\kappa,\gamma \rbrace/2\pi=\lbrace 2.24,0.5,2.9\rbrace\,$MHz for the $|F=2, m_F=2\rangle\leftrightarrow |F^\prime=3, m_F^\prime=3\rangle$ transition of the D$_1$ line of \Rba \ \cite{Wilk2007}. 
The cavity is also coupled to the $|F=1, m_F=1\rangle\leftrightarrow |F^\prime=2, m_F^\prime=2\rangle$ transition of the D$_1$ line of \Rbb. 
The coupling strength $g$ is reduced to $2\pi\times2.12\,$MHz for \Rbb \ because of its slightly smaller transition matrix element. The cavity is detuned by $\Delta/2\pi=1.367\,$GHz from the \Rba \ transition and by $-\Delta$ from the \Rbb \ line, leading to identical emission frequencies. This setting is chosen as neighboring transitions are either far-detuned or forbidden.

Using a 3-level master equation approach, we calculate the dynamics of the vSTIRAP process, and confirm that the power envelopes of the photons emitted by the two atomic species can be made almost perfectly indistinguishable by tuning the Rabi frequencies of the two processes \cite{vasilev2010single, vitanov2017stimulated,SupplMat}.
We further find that the efficiency of the processes and the probability of spontaneous emission can be tuned to achieve a workable success probability $P_S=2\times P_{stim}\times(1-P_{stim})\times P_{coll}\times (1-P_{spon})^2$, where $P_{stim}$, $P_{coll}$ and $P_{spon}$ indicate the probabilities for stimulated emission, photon collection and spontaneous emission, respectively. 
We also extract the probability for false-positive detection, (both atoms emit a photon, but only one is detected), $P_F=P_{stim}^2\times P_{coll}\times (1-P_{coll})$, where we assume number-resolving photon detectors \cite{Wilk2007,SPCM}. 
From the numerical calculations we find  the best ratio $P_{stim}/P_{spon}\simeq 3.2$ for $P_{stim}<0.2$. In this regime there is therefore a simple trade-off between success probability and false-positive detection. 
For example, if we assume $P_{coll}=0.4$  and $P_{stim}=0.1$ (Ref.~\cite{Muecke2013}), then $P_S=7.0\,\%$ and $P_F=0.26\,\%$; if $P_{stim}=0.2$, then $P_S=11.2\,\%$ and $P_F=0.95\,\%$. Spontaneous emission is not problematic as such, since it will in all likelihood lead to a loss of the affected atom from the spatial or temporal detection windows.

Ensuring that the two atoms couple in the same way to the cavity mode requires  their  separation  to be  less than the mode waist ($\sim 40 \ \mu$m) in the radial direction and less than the cavity mode Rayleigh length in the longitudinal direction ($\sim$ few mm). This is not a concern for atoms at few $\mu$K temperatures and a free evolution time $t_0\sim 1 \ms$ between the atom preparation and the vSTIRAP pulse.

Because of the different masses of the two atoms, the recoil is different by $ 2.3\%$ for the two species, which results in different paths followed by the particles (this effect is exaggerated in Fig.~\ref{fig:implementation}). 
For $T = 50 \ms$, the maximum  displacement between the two species within one interferometer path is $\simeq 5 \ \mu m$  \footnote{We note further that the paths of the two isotopes can be made near-identical by operating on the D2 line for \Rbb, and e.g. rendering the emitted photons indistinguishable by wavelength conversion}.

%\Finally, detection of the single atoms after a  total free fall time $\sim t_0 + 2T \sim 100 \ms$ can be achieved with a light sheet detector as in Ref.~\cite{Buecker2009}, where a single atom signal to background ratio exceeding $18$ was reported. For our application, two light sheets vertically separated can be used (each resonant with each atom) in order to resolve in time the two atoms.

\label{par:sensitivity}
We conclude by estimating the sensitivity that could be achieved in a WEP test.  
The interferometer fringes can be reconstructed shot after shot by varying the Raman laser relative phase for one species (e.g. before the last beam splitter), allowing to extract  $\Delta\Phi\stxt{WEP}$. 
Assuming a single-atom quantum projection noise limited sensitivity \cite{Parazzoli2012}, the acceleration sensitivity is given by $\sigma\stxt{WEP}\simeq 1/(k_z T^2 \sqrt{N})$, where $N$ is the number of measurements. 
With $N = 10^4$ successful measurements ($10 \mrad$ phase sensitivity) and $T = 50 \ms$, a differential acceleration sensitivity 
$\sim 5\times 10^{-7} \ \text{m.s}^{-2}$ can be reached, corresponding to a potential sensitivity $\sim 5\times 10^{-8}$ on the E\"otv\"os parameter. 
Note that vibration noise is expected to have a negligible effect as it is common to both interferometers (see Eq.~\eqref{eq:final_phi}).
Further measurements can  then be performed independently with one species at a time to extract the values of the gravitational acceleration separately for each species, and to investigate systematic effects \footnote{Isotope-dependent phase shifts must be carefully controlled. For example, the difference in magnetic field response of the two isotopes corresponds to a systematic effect of the order of $500$~mrad for a magnetic field gradient of $1 \ \text{mG.cm}^{-1}$ and $T = 50 \ms$, which must be controlled at the $2\%$ level to achieve an accuracy of $5\times 10^{-8}$ on the Eotvos  parameter.}.

The effect of entanglement on the free-fall can thus be directly assessed by comparing the  differential gravity obtained with the entangled atoms ($g^\ata-g^\atb$ in $\Delta\Phi\stxt{WEP}$) to that obtained with the classically independent atoms ($g^\ata$ and $g^\atb$ measured independently).

%\paragraph{Conclusion.}
\label{par:conclusion}
Current WEP experiments  performed with cold atoms consist of a differential measurement between two classically independent atom  interferometers \cite{Tarallo2014,Schlippert2014,Zhou2015,Bonnin2015,Duan2016,Rosi2017}.
They explore the validity of the WEP in a different regime than experiments involving macroscopic objects, because the measurement principle involves (single-)atom interference (i.e. a delocalized test mass), and therefore rely on superpositions of quantum degrees of freedom.
For example, the  recent result reported in \cite{Rosi2017} uses an atom in an incoherent superposition of two internal energy states separated by $\sim 30 \ \mu$eV, allowing to probe new possible WEP violations \cite{Zych2015}. 
These experiments probe the WEP with microscopic masses, and should be tested separately to the macroscopic case.
However, all  WEP tests so far compared the gravity acceleration between two classically independent proof masses.
Our proposal makes a conceptual stride beyond previous works, by enforcing quantum entanglement between two  atomic species of different mass ($\sim 2$~GeV energy difference), allowing to probe directly the effect of entanglement on the free fall.
%This is clear from Eq.~\eqref{eq:final_phi}, which shows that the difference in the gravity acceleration between the two atoms is measured coherently in $\Delta\Phi\stxt{WEP}$ as a result of two-particle interferometry \cite{Horne1989,Rarity1990}. 
Specifically, our  scheme could for example be used to assess the quantum formulation of the WEP presented in \cite{Zych2015} at the scale of $2$~GeV \footnote{More precisely, our scheme could be used to constrain the non-diagonal elements of the dimensionless operator $\hat{\eta}=\hat{I}_{int}-\hat{M}_g\hat{M}_i^{-1}$ defined in \cite{Zych2015}, and constrained at the level of $5\times 10^{-8}$ for a $\sim 30 \ \mu$eV energy difference in \cite{Rosi2017}.}.

To the best of our knowledge, there is currently no theoretical model which addresses the question whether or not the presence of entanglement in a system would  result in a violation of the WEP at a given level of accuracy. In general, WEP tests involving  new types of physical objects (compared to macroscopic  masses), such as matter-waves or anti-matter, are motivated by the qualitatively different nature of the involved proof masses rather than by a consensual  theory  predicting a violation in such systems. Our proposal follows this approach by aiming for a test of a foundational principle of Physics with a qualitatively new system not considered before  \cite{Aspelmeyer2017}.

Beyond a conceptually new type of WEP test, our proposal can be used for a test of Bell's inequalities with  free falling massive particles  of different  species.  
Following Ref.~\cite{Dussarrat2017}, a correlation coefficient $E$  can be formed from the measurement of the four  joint probabilities associated to the four modes  appearing in Eq.~\eqref{eq:output_state}. It  reads $E=V\cos(\Delta\Phi) \simeq V\cos\left[k_z T^2 (g_z^\mathcal{A} - g_z^\mathcal{B})\right]$ and can be interpreted as a measure for a Bell test in the presence of gravity.

%It could also be used to  shed new light on the resolution of the Compton-clock debate (\cite{Wolf2011} and references therein). 
%Beyond tests of the WEP, the proposed entangled state of two atomic species represents an intriguing starting point for studies of fundamental physical phenomena.

%\paragraph{Acknowledgments.}
\begin{acknowledgements}
\label{par:acknowledgments}
We thank P. Wolf, C. Garrido Alzar, Y. Sortais, A. Landragin, A. Browaeys, F. Pereira Dos Santos, J.-P. Uzan, and \v{C}. Brukner for fruitful discussions.
\end{acknowledgements}

\bibliographystyle{apsrev4-1}
\bibliography{entangled_interferometer}

\begin{thebibliography}{52}
\bibitem [{foo()}]{footnoteDoppler}%
  \BibitemOpen
  \href@noop {} {}\bibinfo {note} {We neglect here the Doppler effect
  originating from the velocity of the atom in the direction of the
  beams.}\BibitemShut {Stop}%
\end{thebibliography}

% *****************************************
% *****************************************
% *****************************************

\onecolumngrid
\appendix
\newpage

\section{SUPPLEMENTAL MATERIAL}
\textit{In this supplementary material, we first describe the simulation of the photon emission dynamics for the atoms in the cavity, which is used to estimate the probability of success and of false-positive detection events.
We then discuss  peculiar features related to the center-of-mass of the two-entangled-species interferometer when compared to conventional dual-species atom interferometers in which the center-of-mass of the two species are  independent.}

\section{Simulation of the VSTIRAP dynamics}

\subsection{Model}
The single photon which is detected when the $^{87}$Rb and  $^{85}$Rb atoms are in the cavity is created by a vacuum-stimulated Raman adiabatic passage (vSTIRAP). To estimate the performance of the process under the particular conditions of our work, we simulate the vSTIRAP dynamics of the two atoms using a master equation approach. We follow the treatment given in pioneering theoretical \cite{Law1997, Kuhn1999} and experimental work \cite{Kuhn2002, Keller2004}. The method is related to coherent population transfer and electromagnetically induced transparency \cite{Bergmann1998,Fleischhauer2005}.

The coupling of an atomic transition between two levels $i\leftrightarrow j$ to the cavity is given by
\begin{equation}
g_{ij}(z)=\sqrt{\frac{2\pi\mu_{ij}^2}{2\hbar \lambda \epsilon_0 V}}f(z),
\end{equation}
where $\mu_{ij}$ is the relevant transition dipole matrix element, $V$ is the mode volume of the cavity and $f(z)$ is the dependence of the vacuum electric field of the cavity on the vertical position. We assume that the atoms fall through the central portion of the cavity mode, such that the longitudinal and transversal variation of the mode strength is negligible on the scale of the distribution of the atoms' trajectories.

We now describe the evolution of the electronic state of the atoms, starting from an initial state $|S\rangle$ and ending in the final state $|F\rangle$, which are stable ground states of the atoms. Electronic excitation is reduced by introducing a detuning from the relevant transition frequency $\omega_{ij}$. 
The Hamiltonian for a single atom in the cavity is then given by
\begin{equation}
H/\hbar=\Delta_C a^\dagger a+\Delta_P |S\rangle \langle S|+\Omega(t)\left(\sigma_{SE}^\dagger + \sigma_{SE}\right)+g_{FE}(t)\left(\sigma_{FE}^\dagger a + \sigma_{FE} a^\dagger \right),
\end{equation}
where $\sigma_{ij}=|i\rangle\langle j|$ is the atomic inversion operator, and the detuning values for the classical pump field and the cavity are given by $\Delta_{P}=\omega_{P}-\omega_{SE}$ and $\Delta_{C}=\omega_{C}-\omega_{FE}$. The creation and annihilation operators $a^\dagger$ and $a$ relate to the cavity photon occupation.\\
The evolution of the system over time, including spontaneous emission of the atomic excited states and the transmission of generated photons through the cavity mirrors, is described by the master equation
\begin{equation}
\frac{\partial \rho}{\partial t}
=-\frac{i}{\hbar}\left[H,\rho\right]+D(\kappa,a)+D(\gamma_{ES},\sigma_{SE})+D(\gamma_{EF},\sigma_{FE}).
\label{mEq}\end{equation}
The decay evolution term for an operator $o$ and a decay rate $Y$ is given by $D(Y,o)=Y\left( 2 o\rho o^\dagger - o^\dagger o\rho -\rho o^\dagger o \right)$. The amplitude decay rates $\kappa$, $\gamma_{ES}$ and $\gamma_{EF}$ relate to the cavity field, atomic spontaneous decay to the initial state, and decay to the final state, respectively. Assuming that the cavity field decays only by transmission through the output mirror, the probability of photon generation efficiency is given by 
\begin{equation}
P_{stim}=2\kappa\int_0^{\infty}Tr\left\lbrace a^\dagger a \rho \right\rbrace dt.
\end{equation}
Similarly, the probability of an atom undergoing spontaneous emission is given by 
\begin{equation}
P_{spont}=2\left(\gamma_{ES}+\gamma_{EF}\right)
\int_0^{\infty}Tr\left\lbrace |E\rangle \langle E| \rho \right\rbrace dt.
\end{equation}

\subsection{Transitions and parameters}\label{subsec:setup}

\begin{figure}[t!]
\centering
\includegraphics[width=\columnwidth]{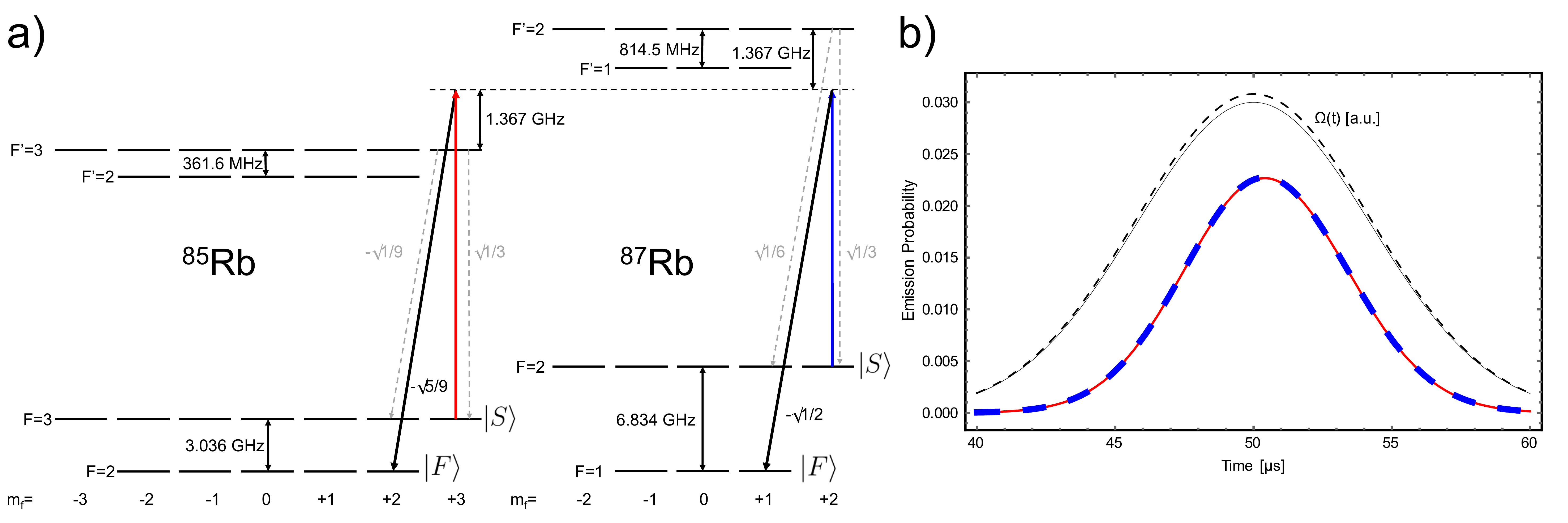}
\caption{a) Schematic representation of the energy levels involved in the D$_1$ transitions in \Rba \ and \Rbb \ (not to scale). The dashed horizontal line indicates the virtual level through which the Raman transitions are operated. The dashed gray lines are undesired spontaneous decay channels. The relevant Clebsch-Gordan coefficients are shown. The black arrows indicate the transitions resonant to the cavity, while the red and  blue arrows denote the pump transitions. 
b) Example trace of the photon emission probability  as a function of time for the two isotopes (plain red line  for \Rba,  dashed blue line for \Rbb), showing near-identical emission profiles.
The black lines show the pump fields $\Omega(t)$ for \Rba \ and \Rbb.  
}\label{fig:specPic}
\end{figure}

We simplify our simulation to three levels by compounding the two decay channels occurring in transition $E \rightarrow S$ into a single effective decay back into state $|S\rangle$ (see Fig. \ref{fig:specPic} a), dashed gray lines). The two decay channels $E \rightarrow S$ and $E \rightarrow F$  are scaled with the square of the Clebsch-Gordan coefficients.
In \Rbb, $\gamma_{ES}=\gamma_{EF}$ (since $1/3 + 1/6 = 1/2$), and in \Rba, $\gamma_{ES}=4\gamma_{EF}/5$ (since $1/3 + 1/9=4/9=4/5\times 5/9$). 
Note that while $\gamma_{EF}$ only results in the loss of an atom, $\gamma_{ES}$ can result in the detection of a photon and concurrent loss of the atom. Therefore the experimental data will have to be post-selected on sequences in which one photon and both atoms are detected within the designated spatial and temporal windows.

We envisage a cavity with similar parameters as used in \cite{Wilk2007}, but in a ring geometry, with  coupling strength, field amplitude decay and atomic decay rates $\lbrace g,\kappa,\gamma \rbrace/2\pi=\lbrace 2.24,0.5,2.9 \rbrace\,$MHz for the $|F=2, m_F=2\rangle\leftrightarrow |F^\prime=3, m_F^\prime=3\rangle$ transition of the D$_1$ line of the \Rba \ atoms.
The $|F=1, m_F=1\rangle\leftrightarrow |F^\prime=2, m_F^\prime=2\rangle$ transition of the D$_1$ line of \Rbb \ also couples to the cavity. 
The \Rba \ and \Rbb \ atoms enter the cavity in the states $|F=3, m_F=3\rangle$ and $|F=2, m_F=2\rangle$, respectively, and are driven with individual, $\pi$-polarized pump beams. 
The coupling strength $g_{max}$ for \Rbb \ is reduced to $2\pi\times2.12\,$MHz because of its slightly smaller transition matrix element. This reduction can be compensated for by a small increase of the driving amplitude, in order to create indistinguishable wavepackets. The cavity is detuned by $\Delta/2\pi=1.367\,$GHz from the \Rba \ transition and by $-\Delta$ from the \Rbb \ line, leading to identical emission frequencies. This setting is chosen as neighboring transitions are either far-detuned or forbidden (See Fig. \ref{fig:specPic} a)).
\subsection{Emission}\label{subsec:emission}
\begin{figure}[t!]
\centering
\includegraphics[width=1.\columnwidth]{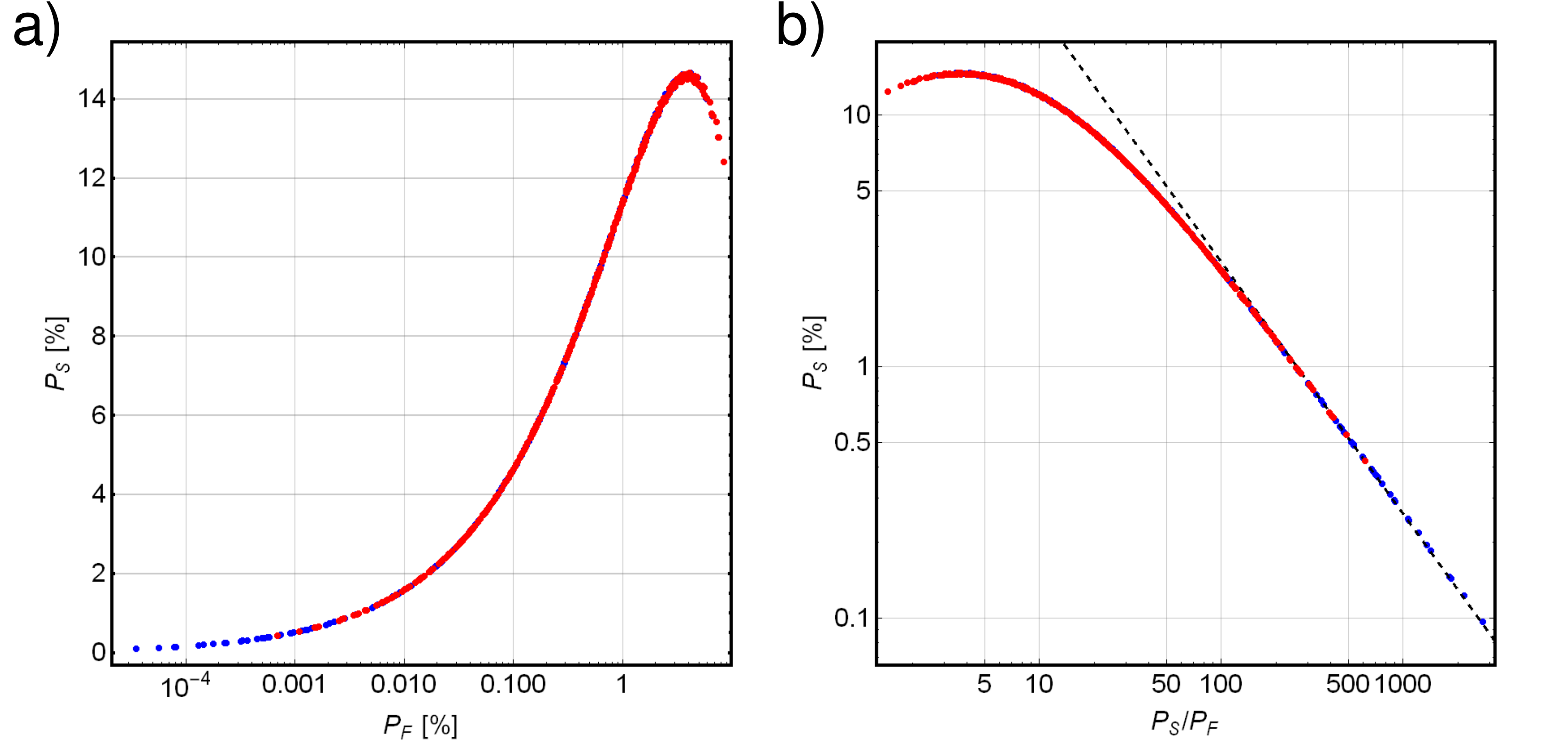}
\caption{a) Log-linear plot of the probability of success $P_S$ versus the probability of a false-positive event $P_F$. \Rbb \ and \Rba \ (blue and red dots) show near-identical behaviour. b) Logarithmic plot of the ratio of the two probabilities. For comparison, the dashed black line shows a linear dependence.}\label{fig:success}
\end{figure}
%The atoms transit through the cavity in free fall, after an acceleration of $\sim 100\,$ms, and have therefore attained a velocity of $v\sim 1\,$m/s. 

The cavity mode is assumed to have a waist of $w_0=40\,\mu$m. The atoms are prepared in the cavity (e.g. by laser cooling in optical tweezers) and are dropped from their  trap shortly before the beginning of the driving fields, which occur on a  time-scale $t_0 \lesssim 1$~ms after the drop. 
We therefore assume that the atoms experience a constant coupling $g_{max}$ to the cavity mode. 
%by $g(t)=g_{max}e^{-((z_0-v t)/w_0)^2}$. 
The coupling to the cavity, together with a time-dependent excitation pulse amplitude, controls the shape of the photon pulse emitted by the atom-cavity system. 
In order to properly control the emission characteristics,   a number of limiting factors need to be considered. The cavity linewidth sets a suitable lower bound for the emission pulse bandwidth, and also for the Zeeman splitting required to lift the degeneracy between allowed two-photon transitions in the electronic state manifolds of the two atomic species. 
%Conversely, the transit time sets an upper limit for the duration of the photon envelope. 
We therefore examine Gaussian pulse envelopes of the type $\Omega(t)=\Omega_{max}e^{-((t-t_0)/\tau)^2}$ with durations between $\tau=2\,\mu$s and $\tau=10\,\mu$s, for which the aforementioned limits are not expected to have a significant effect. 
We numerically explored the parameter space by varying the pump field Rabi frequency $\Omega_{max}$ (in the range of tens of MHz) and the pulse duration $\tau$, and computed the probability of stimulated emission, $P_{stim}$, and that of spontaneous emission, $P_{spon}$.
%The stimulated emission probability was found to be maximum for pulses that occur when the atom is at the maximum of the cavity mode, as expected. 
 The numerical exploration indicates that the ratio $P_{stim}/P_{spon}\simeq 3.2$, which limits the probability of success.

Fig.~\ref{fig:success} a) shows the result of the numerical exploration (each dot corresponds to a value of the tuplet $\{\Omega_{max},\tau,t_0\}$) by displaying the success probability,  $P_S=2\times P_{stim}\times(1-P_{stim})\times P_{coll}\times (1-P_{spon})^2$, versus the false-positive probability, $P_F=P_{stim}^2\times P_{coll}\times (1-P_{coll})$. The figure shows that $P_S > 14\,$\% can be achieved for these parameters. We emphasize that this value is limited purely by technological factors, such as the cavity finesse and mode field diameter. Fig. \ref{fig:success} b) shows the ratio of $P_S/P_F$, underlining the trade-off between the rate of successful and false-positive events. 
As exemplary values, $P_{stim}=10.4\,$\% leads to $P_S=7.0\,$\% and $P_F=0.26\,$\%, and $P_{stim}=20.0\,$\% leads to $P_S=11.2\,$\% and $P_F=0.95\,$\%.
These values for $P_F$ and $P_S$  rely on a modest assumption for the photon detection efficiency, since modern superconducting detectors approach unity efficiency. The value of $P_{coll}=0.4$ assumes large path losses and scattering or absorption in the cavity output mirror. This value can also be improved by purely technical means.

\section{Center-of-mass of the two-species interferometer}

The usual approach for describing the phase difference in an atom interferometer is  the path integral formulation of quantum mechanics (see, e.g. Ref.~\cite{Storey1994}). 
Following a semi-classical approach, it has been shown that the phase shift in a symmetric 3 light-pulse atom interferometer originates from the relative phase of the Raman lasers which is imprinted on the atomic wavefunction at the atom-laser interaction times \cite{Wolf2011}, such that the interferometer acts as an accelerometer \footnote{In a representation-free, full-quantum description, it has been shown that the interferometric phase originates from  a product of non-commuting unitary operators which reflects the acceleration of the atom in the laser frame \cite{Schleich2013,Schleich2013a}.}. 
We followed this semi-classical approach in our work: we evaluated the phase shifts imprinted on each atomic species by following the classical trajectories of the two atoms in the arms of the interferometer. 

In single-atom interferometers, according to the midpoint theorem \cite{Antoine2003}, the interferometric phase can be computed by calculating the classical trajectory of the center-of-mass (COM) between the two arms of the interferometer, which is not a populated trajectory. 
In dual-species single-atom interferometers, the   trajectories of the COM associated with the two species are physically  separated and   independent. 
In our proposal, remarkably, the COM associated with the two possible states forming the superposition of Eq.(1) of the main text follow different trajectories. Moreover, the two COMs associated to the two species are not equidistant from the total COM of the two-atom entangled state, within which interference occurs. 
In this section, we analyze  this non-local feature specific to the two-atom interferometer  in more details.

Consider  the state $|\ata,\hbar \vec{k}_\ata; \atb, \vec{0}\rangle$ where atom $\ata$ recoils with a velocity $\hbar \vec{k}_\ata/m_\ata$ and atom $\atb$ is left unperturbed. 
We call $COM_1(t)$ the center-of-mass trajectory for that state in the vertical ($z$) direction \footnote{We recall that the wavevectors in the $x$ (cavity) direction are the same for the two species within the cavity mode linewidth. Only the wavevectors associated with the pump beams in the $z$ direction are different.}. Conversely, we call $COM_2(t)$ the center-of-mass trajectory  corresponding to the state $|\ata, \vec{0}; \atb, \hbar \vec{k}_\atb\rangle$, where $\atb$ recoils with the velocity  $\hbar \vec{k}_\atb/m_\atb$. 
%The different paths are shown in Fig.~\ref{fig:illustration_COM_difference}; by convention, we call $(AB)-(BC)$ the path corresponding to a recoil at the first pulse and $(AD)-(DC)$ the  path which is unperturbed at the first pulse. 
If we concentrate on the first part of the interferometer ($0\leq t \leq  T$), the trajectories are given by 
\begin{equation}
COM_1(t)  = \frac{\hbar k_{\ata}}{m_{tot}} t \ \ , \ \ COM_2(t)  = \frac{\hbar k_{\atb}}{m_{tot}} t,
\label{eq:COM_state}
\end{equation}
with $m_{tot}=m_{\ata}+m_{\atb}$.
%Consider  the first state $|$\Rba$,\hbar\vec{k}_{85}; $ \ \Rbb$,\vec{0}\rangle$ of Eq.(1) of the main text, where \Rba \ recoils with a velocity $v_{85}=\hbar k_{85}/m_{85}$ and \Rbb \ is left unperturbed. We call $COM_1(t)$ the trajectory of the COM for that state. Conversely, we call $COM_2(t)$ the trajectory of the COM corresponding to the state where \Rbb \ recoils and \Rba \ is left unperturbed. The corresponding trajectories are, for $0<t<T$:
%\begin{equation}
%COM_1(t)  = \frac{\hbar k_{85}}{m_{85}+m_{87}} t \ \ , \ \ COM_2(t)  = \frac{\hbar k_{87}}{m_{85}+m_{87}} t.
%\end{equation}
The trajectory of the total center-of-mass of the entangled state  is:
\begin{equation}
COM_{tot}(t)  = \frac{1}{2}\left[COM_1(t)+COM_2(t)\right] = \frac{\hbar k_{\ata} + \hbar k_{\atb}}{2m_{tot}} t,
\label{eq:COM_tot}
\end{equation}
for all times $0\leq t \leq 2T$.
The center-of-mass trajectories for each atomic species are   $COM_\alpha(t)=\hbar k_\alpha/2 m_\alpha \times t$, where $\alpha = \ata,\atb$. 
The relative trajectories between the center-of-mass of each species and the total center-of-mass are:
\begin{eqnarray}
COM_{\ata}(t)-COM_{tot}(t) &  = & \frac{\hbar k_\atb \times t}{2 m_{tot}} \left(\frac{m_\atb k_\ata}{m_\ata k_\atb} -1 \right) \nonumber  \\
COM_{\atb}(t)-COM_{tot}(t) &  = & \frac{\hbar k_\ata \times t}{2 m_{tot}} \left(\frac{m_\ata k_\atb}{m_\atb k_\ata} -1 \right).
\label{eq:distance_COM}
\end{eqnarray}
The  center-of-mass trajectories associated to the two species are, in general, not equidistant from the total center-of-mass of the full state. 

%This effect is illustrated in Fig.~\ref{fig:distance_to_COM}, where we plot $d_{85,87}(t)=COM_{85,87}(t)-COM_{tot}(t)$ for \Rba \ and \Rbb \  interrogated on the $D_1$ line, as considered in the main text of our article.

\begin{figure}[t!]
\centering
\includegraphics[width=1.\columnwidth]{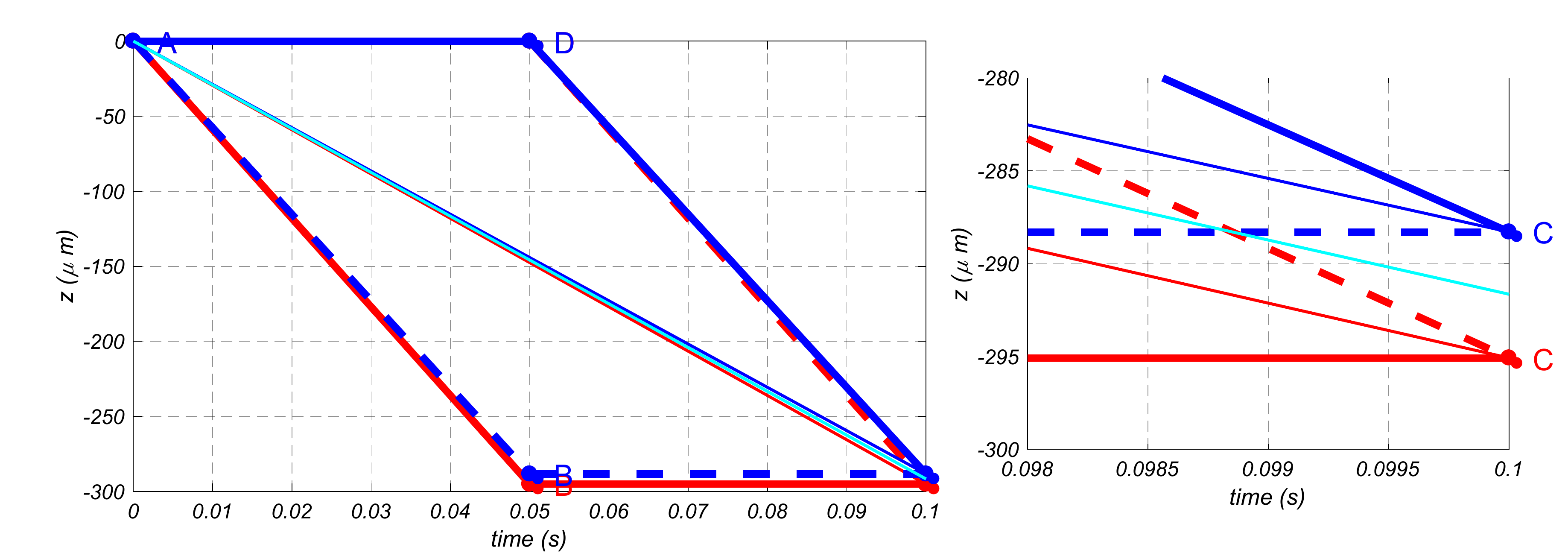}
\caption{Illustration of the splitting of the center-of-mass trajectory  of each isotope (blue and red lines) from the trajectory of the total center-of-mass of the two-atom entangled state (cyan line). The right panel is a zoom to the final two-atom interferometer recombination point.
%Distance from the center of mass  of each isotope to the total center-of-mass of the two-atom entangled state, $COM_{tot}$, as a function of time. 
%The plain red line is for \Rba, the dashed blue line for \Rbb. 
The time between the light pulses is $T=50$~ms.  
%Bottom panel: relative distance of the isotope to the COM. Because of the difference in recoil between the two species ($\sim 2\%$), the heaviest atom (here \Rbb) is closer to the COM than the lightest species (\Rba). 
Gravity is set to zero to emphasize the recoil effect.
}
\label{fig:distance_to_COM}
\end{figure}

With our choice of atomic species and transition lines ($D_1$ for both isotopes), the relative difference between the two wavevectors is small, $(k_{85}-k_{87})/k_{85}\sim 10^{-5}$, but the relative mass difference is comparatively larger, $(m_{87}-m_{85})/m_{87}\simeq 0.023$. From Eqs.~\eqref{eq:distance_COM}, the displacement between the center-of-mass of each atom and $COM_{tot}$ is macroscopic ($ 3.43$ and $-3.35 \ \mu$m for \Rba \ and \Rbb, respectively), as mentioned in the main text (Ref.~[48]). This effect is illustrated in Fig.~\ref{fig:distance_to_COM}.

In dual-species single-atom interferometers, the interference occurs independently within the center-of-mass of each species, corresponding to the two C points in Fig.~\ref{fig:distance_to_COM}. 
The displacement between the center-of-mass of the atoms has therefore no fundamental role, but is of technical concern in WEP tests because of gravity gradients.
In our two-atom interferometer, the interference occurs within the total center-of-mass of the two-atom state (end point of the cyan line in Fig.~\ref{fig:distance_to_COM}), which is separated from the classical recombination points associated to each species.

Such peculiar non local effects could be magnified by using different transitions in both atoms, such as the $D_2$ (780 nm) and $D_1$ (795 nm)  lines for \Rba \ and \Rbb \ respectively, or two different atomic species such as $^{85}$Rb and $^{133}$Cs (780 nm and 852 nm for $D_2$ lines). In such scenarios, indistinguishability of the emitted photons can  be engineered by means of wavelength conversion. %Such configurations are illustrated in Fig.~\ref{fig:illustration_COM_difference}.

\section{References}

%\bibliographystyle{apsrev4-1}
%\bibliography{../../entangled_interferometer}

\end{document}